 \documentclass[preprint,review,12pt]{elsarticle}

\usepackage{amssymb}
\usepackage{float} 
\usepackage{multirow}
\usepackage{physics}
\usepackage{hyperref}
\hypersetup{
    colorlinks=true,
    linkcolor=blue,
    filecolor=magenta,      
    urlcolor=cyan,
}

\usepackage[flushleft]{threeparttable}
\usepackage{tabu}
\usepackage{longtable}
\usepackage{booktabs}
\usepackage[dvipsnames]{xcolor}

\journal{}

\begin{document}
\begin{frontmatter}

\title{Multi-input model uncertainty analysis for long-range wind farm noise predictions}

\author{Phuc D. Nguyen\corref{cor1}\fnref{label1}}
\author{Kristy L. Hansen\fnref{label1}}
\author{Branko Zajamsek\fnref{label2}}
\author{Peter Catcheside\fnref{label2}}
\author{Colin H. Hansen\fnref{label3}}

\cortext[cor1]{ducphuc.nguyen@flinders.edu.au}
\address[label1]{College of Science and Engineering,  Flinders University, Adelaide, SA 5042, Australia}

\address[label2]{Adelaide Institute for Sleep Health,  Flinders University, Adelaide, SA 5042, Australia}

\address[label3]{School of Mechanical Engineering, University of Adelaide, Adelaide, SA 5005, Australia}

\begin{abstract}

One of the major sources of uncertainty in predictions of wind farm noise (WFN) reflect parametric and model structure uncertainty. The model structure uncertainty is a systematic uncertainty, which relates to uncertainty about the appropriate mathematical structure of the models. Here we quantified the model structure uncertainty in predicting WFN arising from multi-input models, including nine ground impedance and four wind speed profile models. We used a numerical ray tracing sound propagation model for predicting the noise level at different receivers. We found that variations between different ground impedance models and wind speed profile models were significant sources of uncertainty, and that these sources contributed to predicted noise level differences in  excess of 10 dBA at distances greater than 3.5 km. We also found that differences between atmospheric vertical wind speed profile models were the main source of uncertainty in predicting WFN at long-range distances. When predicting WFN, it is important to acknowledge variability associated with different models as this contributes to the uncertainty of the predicted values.

\end{abstract}




\end{frontmatter}


\section{Introduction}

A crucial step to reduce potential impacts of noise on humans is to maximise the accuracy of noise prediction models. Accurate noise predictions during the planning stage of a new project can reduce the chance of possible exceedances of relevant allowable limits during the operational stage \cite{hansen2018wind}. Recent advances in our knowledge and computational resources have allowed for the development of complex sound propagation models \cite{salomons2012computational,ostashev2015acoustics,jensen2011computational}, to help account a variety of complex input parameters such as atmospheric and topographical conditions. Despite recent advances in modelling methods, predicting outdoor noise levels still remains challenging and high levels of uncertainty remain \cite{wilson2007characterization,van2018variability,wilson2014description}. 

Uncertainty in predicting noise levels can be attributed to differences in specifying model parameters (parameter uncertainty) and the choice of prediction model (model uncertainty) \cite{ostashev2015acoustics}.  For example, parameter uncertainty includes uncertainty in ground flow resistivity due to variations in the ground composition between source and receiver, terrain profile and vertical atmospheric sound speed profile, which can also vary significantly between source and receiver \cite{attenborough1995benchmark,kayser2020environmental,wilson2014description}. Model uncertainty includes physical and mathematical assumptions and numerical approximations \cite{ostashev2015acoustics}, which are different for different models. Recent progress has been made to quantify parameter uncertainty such as uncertain geometries \cite{parry2020investigating}, atmospheric turbulence \cite{hormeyer2019prediction}, meteorological states and ground properties \cite{van2018variability,wilson2007characterization,leroy2010uncertainty}.  In particular, by modelling the distribution of input parameters such as the atmospheric vertical wind speed profile, flow resistivity and porosity of the ground, derived from experimental data, \citet{van2018variability} found that 95\% confidence interval around predicted noise levels can be up to 10 dBA, even for short range propagation ($<$ 250 m). Through investigating the uncertainty associated with source and receiver positions at short-range ($<$ 200 m), \citet{parry2020investigating} found that the statistical distributions of propagation attenuation spectra were strongly affected by location uncertainties associated with these positions. The sensitivity of the parabolic equation model (PE), a widely-used model for outdoor noise propagation, has also been extensively investigated \cite{pettit2007proper,kayser2019sensitivity,kayser2020environmental}.  \citet{kayser2020environmental} found that the most sensitive parameters include the atmospheric vertical wind speed profile coefficient, the angle between the wind and the source-receiver direction and the flow resistivity of the ground.

If the most sensitive parameters such as atmospheric vertical wind speed profile coefficient and flow resistivity are well-determined, then the uncertainty would be expected to be minimised. However, these parameters also need to be used as inputs to impedance \cite{attenborough2011outdoor} and atmospheric vertical wind speed profile models \cite{gualtieri2019comprehensive} (hereafter referred to as input models) to estimate ground reflection coefficients and the atmospheric vertical sound speed profile, which can then be used in sound propagation models such as PE or ray tracing \cite{ostashev2015acoustics}. Consequently, uncertainty in the ground impedance or wind speed profile models will contribute to uncertainty in the final results provided by the model. Although model uncertainty is considered extensively in other fields such as crop prediction \cite{asseng2013uncertainty} or climate change \cite{tebaldi2005quantifying}, it is largely overlooked in outdoor noise level prediction models. 

The purpose of this study was to quantify and characterise the uncertainty in outdoor noise level predictions from input model uncertainty. Specifically, we first sought to ascertain whether the uncertainty associated with atmospheric vertical wind speed profile models and ground impedance models is non-negligible for a simplified controlled case with flat terrain and fixed input values. We then used interaction and partition analysis to investigate the characteristics of these uncertainty sources over long-range propagation.

\section{Methods}
\subsection{Study design}

\begin{figure}[H]
\begin{center}
\includegraphics[width=14cm]{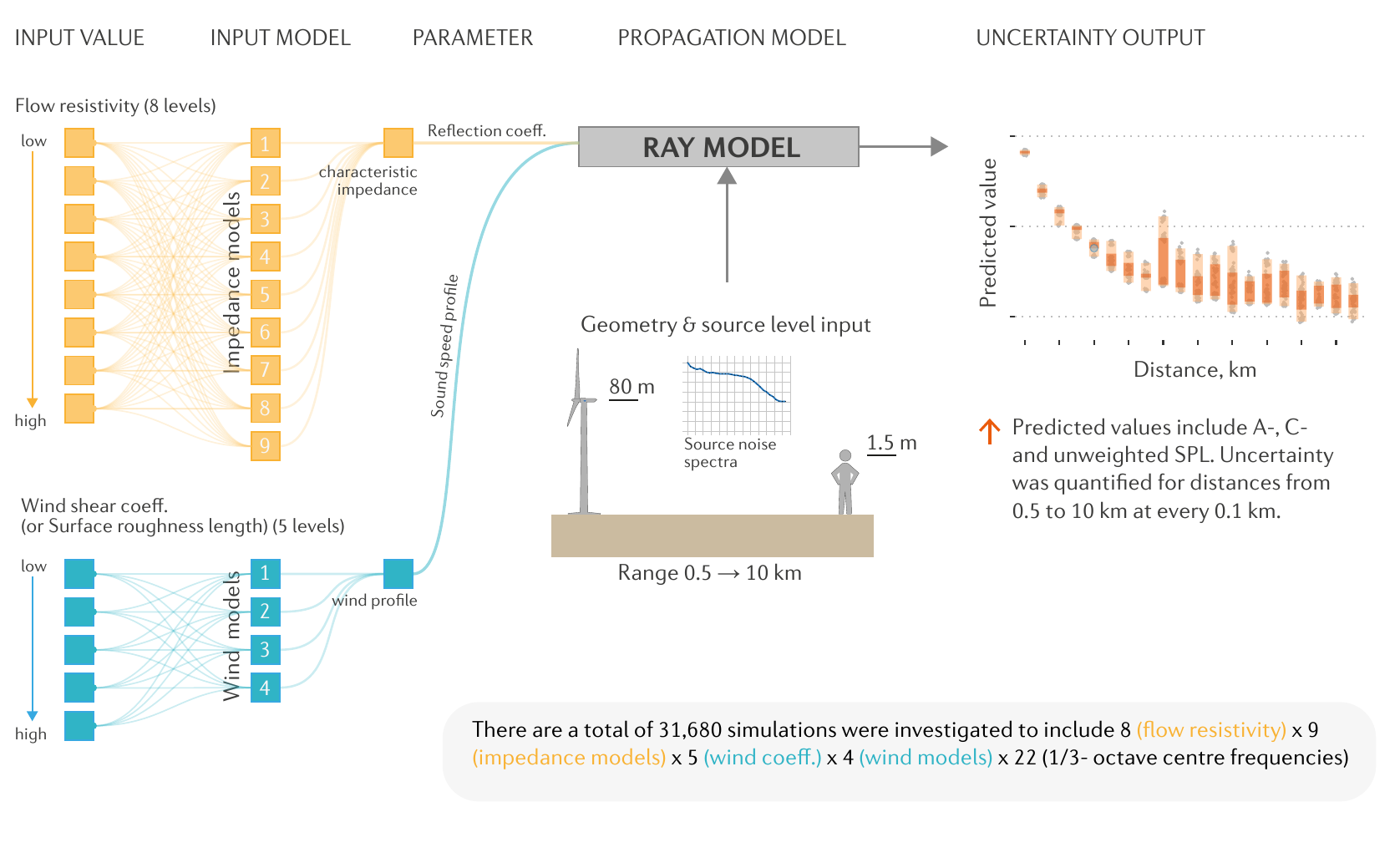}
\end{center}
\caption{{Uncertainty propagation pipeline.}}
\label{fig:FIG1}
\end{figure}

The study design is summarised in \autoref{fig:FIG1}. Eight ground flow resistivity values and five wind shear coefficients (or surface roughness lengths) were input into nine impedance models and four atmospheric vertical wind speed profile models, respectively. The impedance model outputs (i.e., characteristic impedance values) and atmospheric vertical wind speed profile model outputs were then used to calculate plane-wave reflection coefficients and atmospheric vertical sound speed profiles. These parameters were then input into a numerical ray tracing sound propagation model implemented using BELLHOP \cite{porter2011bellhop}. Other input parameters were fixed, including source location and height (80 m), receiver height (1.5 m) and the wind turbine source noise spectrum (See Supplementary Fig. S1 for details). The uncertainties in predictions of A-, C- and unweighted sound pressure levels (SPLs) were quantified at every 0.1 km between 0.5 and 10 km from the turbine.

Some input parameters were simplified to make the problem tractable and to improve computational efficiency whilst retaining conventional model complexity needed to address the primary study aims. Specifically, the terrain profile was assumed to be flat, which is commonly used by practitioners in outdoor wind farm noise predictions \cite{Hansen2017}. This assumption also eliminates additional uncertainty associated with the effects of more complex terrain. Although the PE model has been recommended for outdoor noise propagation \cite{ostashev2015acoustics}, the numerical ray tracing model is more efficient, which makes it more suitable for uncertainty quantification where computational speed is critical \cite{jensen2011computational}. Also, compared to experimental data measured at wind farms in Europe, ray tracing model predictions agree well with the measurements in flat or smooth even terrain \cite{prospathopoulos2007application}. Three levels of ground flow resistivity were used, representing ground surface types from very soft, normal uncompacted ground to compacted dense ground \cite{bies2017engineering}. Also, three levels of the wind shear coefficient (or surface roughness lengths) were investigated to cover the typical expected range of meteorological conditions. Only uncertainty in predicting SPL in the downwind direction was investigated. Fixed input parameters of source and receiver heights were chosen based on the average hub height and the height of the human ear above the ground, respectively. The Suzlon S88 2.1 MW wind turbine noise spectra were obtained from experimental data \cite{Andrew2006}. Finally, overall noise level weightings including  A-, C- and unweighted SPLs were determined because they are the most common metrics used for outdoor noise predictions \cite{Hansen2017,bies2017engineering}. Details regarding the calculation of A-, C- and unweighted SPLs from source noise levels, with propagation loss included, are shown in the Supplementary algorithm 1.

\subsection{Ground impedance models}

Nine commonly-used impedance models were implemented as listed in Table \ref{tab:table1}. These models were selected based on a comprehensive review paper \cite{attenborough2011outdoor}, outdoor sound propagation textbooks \cite{ostashev2015acoustics,salomons2012computational}, and recent publications \cite{horoshenkov2019three}. Only impedance models with four or less input parameters were included in the analysis. Detailed equations can be found in the original papers or in Supplementary table S1. Codes for implementing these models were written using the Julia computer programming language (\href{http://www.r-project.org/}{https://julialang.org/}).

\begin{table}[tb]
\centering
\caption{\label{tab:table1}Impedance models and required input parameters.}

\begin{tabular}{p{8cm} p{2.5cm} p{1.5cm}}

\hline\hline
Model & Input parameter* & Reference\\
\hline
Imp1: Delany \& Bazley model & $\sigma$ & \cite{attenborough2011outdoor,delany1970acoustical}\\
Imp2: Variable porosity model & $\sigma$, $\alpha$ & \cite{attenborough2015outdoor}\\

Imp3: Wilson relaxation model & $\sigma$, $\Omega$ & \cite{wilson1997simple,ostashev2015acoustics}\\

Imp4: Zwikker \& Kosten model & $\sigma$, $\Omega$, $q^2$ & \cite{attenborough2011outdoor,zwikker1949sound}\\

Imp5: Taraldesen's model & $\sigma$, $\Omega$, $q^2$ & \cite{taraldsen2005delany}\\

Imp6: Hamet phenomenological model & $\sigma$, $\Omega$, $q^2$ & \cite{berengier1997porous}\\

Imp7: Identical tortuous slit-like pore model  & $\sigma$, $\Omega$, $q^2$ & \cite{attenborough1987acoustic}\\

Imp8: Horoshenkov's three parameter model &  $\Omega$, $\bar{s}$, $\sigma_s $ & \cite{horoshenkov2019three,horoshenkov2016asymptotic}\\

Imp9: Attenborough's four parameter model & $\sigma$, $\Omega$, $q^2$, $s_f$ & \cite{attenborough1987acoustic}\\

\hline\hline
\end{tabular}
\begin{tablenotes}[flushleft]
         \item[]* \footnotesize The input parameters include: flow resistivity $\sigma$, rate of porosity variation with depth $\alpha$, porosity $\Omega$, tortuosity $q^2$, pore shape factor $s_f$, mean pore size $\bar{s}$, and standard deviation in log-normal mean pore size $\sigma_s$.
\end{tablenotes}

\end{table}

\autoref{tab:table2} lists input values for the impedance models. Eight ground flow resistivity values were used which represent eight types of ground: very soft to water surface (See table 5.2 in \cite{bies2017engineering}). Other parameters were chosen by calculating the median value of their typical range (See Supplementary Table S2 for details). The flow resistivity, $\sigma$, was used for all models except model Imp8 \cite{horoshenkov2019three} which uses mean pore size, $\bar{s}$, instead. However, to simplify the analysis, the mean pore size in model Imp8 was estimated from the flow resistivity using Eq. (56) in \cite{horoshenkov2019three}. The flow resistivity was the focus of the present study because it is one of the most important parameters for estimating the ground effect in noise propagation models \cite{kayser2019sensitivity}. Our sensitivity analysis also showed this to be the most sensitive parameter for the impedance model (See Results section).

\begin{table}[H]
\centering
\caption{\label{tab:table2}Input values for impedance models.}

\begin{tabular}{p{6cm} p{3cm}}
\hline\hline
Parameter, unit & Value\\
\hline
Flow resistivity ($\sigma$), KPa s m$^{-2}$ &$\{12.5, 31.5,80, 200,500,2000, 20000, 200000 \}$ \\
Porosity ($\Omega$) & 0.4\\
Tortuosity (q) & 1.4  \\
Porosity variation rate ($\alpha$), m$^{-1}$ & 5.5  \\
Pore shape factor ($s_f$) & 0.75  \\
Standard deviation of mean pore size ($\sigma_s$) & 0.3  \\

\hline\hline
\end{tabular}
\end{table}

The characteristic impedance values were used to calculate the plane-wave reflection coefficient. The plane-wave reflection coefficient was used instead of the spherical-wave reflection coefficient as recommended by Ostashev and Wilson \cite{ostashev2015acoustics}, given that the wavefront geometry assumptions used to derive the spherical-wave coefficient are not applicable when refraction is present (\cite{ostashev2015acoustics}, p. 359). Therefore, the more complicated formula used to model spherical-wave reflection does not necessarily provide better results. The plane-wave reflection coefficient, $R_p(\psi, \omega)$, that depends on the grazing angle $\psi$ and frequency $\omega$ is calculated as follows:

\begin{equation} \label{eq1}
\begin{split}
&R_p(\psi, \omega) = \frac{\sin \psi  -1/Z_c (\omega)}{\sin \psi  +1/Z_c(\omega)} , \\
\end{split}
\end{equation}
where $Z_c(\omega)$  is the normalised acoustic impedance of the ground, which is frequency dependent.

\subsection{Wind extrapolation models}

Four atmospheric vertical wind speed profile models were implemented as listed in \autoref{tab:table3}. These models are commonly used in the wind energy industry, as outlined in a comprehensive review paper by \citet{gualtieri2019comprehensive} and outdoor noise propagation textbooks \cite{ostashev2015acoustics,salomons2012computational}. We also limited our study to general atmospheric vertical wind speed profile models which are applicable to a range of conditions from unstable to stable. The power law and logarithmic law-based models account for greater than 90\% of wind energy applications \cite{gualtieri2019comprehensive}. Calculation details can be found in \cite{gualtieri2019comprehensive} or Supplementary Table S3 and our open-source code as provided in Section 2.6.  

\begin{table}[H]
\centering
\caption{\label{tab:table3}Wind extrapolation models and required input parameters.}

\begin{tabular}{p{7cm} p{4cm}  p{1cm}}

\hline\hline
Model & Input parameter*& Reference\\
\hline
Wind1: Log-linear law model & $z_0$ & \cite{monin1954dimensionless}\\


Wind2: Power law model & $\alpha$  & \cite{hellmann1919bewegung}\\



Wind3: Smedman-H{\"o}gstr{\"o}m and H{\"o}gstr{\"o}m model & $z_0$ & \cite{smedman1978practical}\\

Wind4: Panofsky model & $z_0$  & \cite{panofsky1960diabatic}\\

\hline\hline
\end{tabular}
\begin{tablenotes}[flushleft]
         \item[]* \footnotesize The input parameters include $z_0$- roughness length; $\alpha$- wind shear coefficient.
\end{tablenotes}
\end{table}

There are two required input parameters for these models which include the wind shear coefficient ($\alpha$) and roughness length ($z_o$). These input values are shown in Table \ref{tab:table4}  derived from our one year data set measured at 1.3 km from the nearest wind turbine of a South Australian wind farm.  The five levels of the wind shear coefficient and surface roughness lengths correspond with the 2.5$^{\rm{th}}$, 25$^{\rm{th}}$, 50$^{\rm{th}}$ 25$^{\rm{th}}$ and 97.5$^{\rm{th}}$ percentiles of the distribution of data measured over one year (See Results section).

\begin{table}[H]
\centering
\caption{\label{tab:table4}Input values for wind models.}

\begin{tabular}{p{6cm} p{3cm}}
\hline\hline
Parameter, unit & Value \\
\hline
Wind shear coefficient ($\alpha$) &$\{0.001,0.13,0.21,0.39,0.79\}$ \\
Surface roughness length ($z_0$), m & $\{0.001, 0.002,0.03,0.26, 0.87\}$\\
Reference height ($h_{ref}$), m & 10   \\
Reference wind speed ($v_{ref}$), m/s & 2.7  \\

\hline\hline
\end{tabular}
\end{table}

\subsection{Ray tracing model}

To predict noise levels at receivers, we used a numerical ray tracing method \cite{ostashev2015acoustics}. This method calculates ray paths and their amplitude, allowing the acoustic pressure field at receivers to be estimated. Although the inherent high-frequency approximations \cite{ostashev2015acoustics,jensen2011computational} of ray tracing models lead to somewhat coarse accuracy in the results, ray tracing models are more efficient and thus more suitable for uncertainty quantification than more computationally demanding methods \cite{jensen2011computational}. 

To implement the ray tracing model, we used BELLHOP, a comprehensive open-source ray tracing program, written in FORTRAN by \citet{porter2011bellhop} and widely used for underwater acoustic sound propagation \cite{jensen2011computational}. Bellhop was developed to model underwater acoustic wave guiding between the sea bed and the water-air interface. These typical boundary conditions in Bellhop are different to atmospheric acoustic sound propagation, for which the sound wave travels between the ground surface and the open air space. To transfer Bellhop to atmospheric sound propagation applications, we made some modifications to the Bellhop model. There are several ways to transfer typical boundary conditions in Bellhop to atmospheric sound propagation applications. One of the most convenient features of Bellhop is that it allows users to input the reflection coefficients for the top and bottom boundary conditions. This option makes the adaptation of Bellhop straightforward. Specifically, we used the reflection coefficient $R = 0$ for the top layer to model the open air space boundary as used in \cite{hussain2020parametric}, resulting in the top boundary being assigned as a no reflection boundary. For the ground surface boundary, we used the reflection coefficients calculated using Eq. (\ref{eq1}). Other input parameters such as the sound speed profile and terrain profile can be inputted directly to Bellhop without any major modifications. We also developed a wrapper package in Julia programming language to implement Bellhop. The source code and several example cases were also provided in \href{https://ducphucnguyen.github.io/FreeRay.jl/build/}{https://github.com/ducphucnguyen/FreeRay.jl}.

\begin{figure}[H]
\begin{center}
\includegraphics[width=14cm]{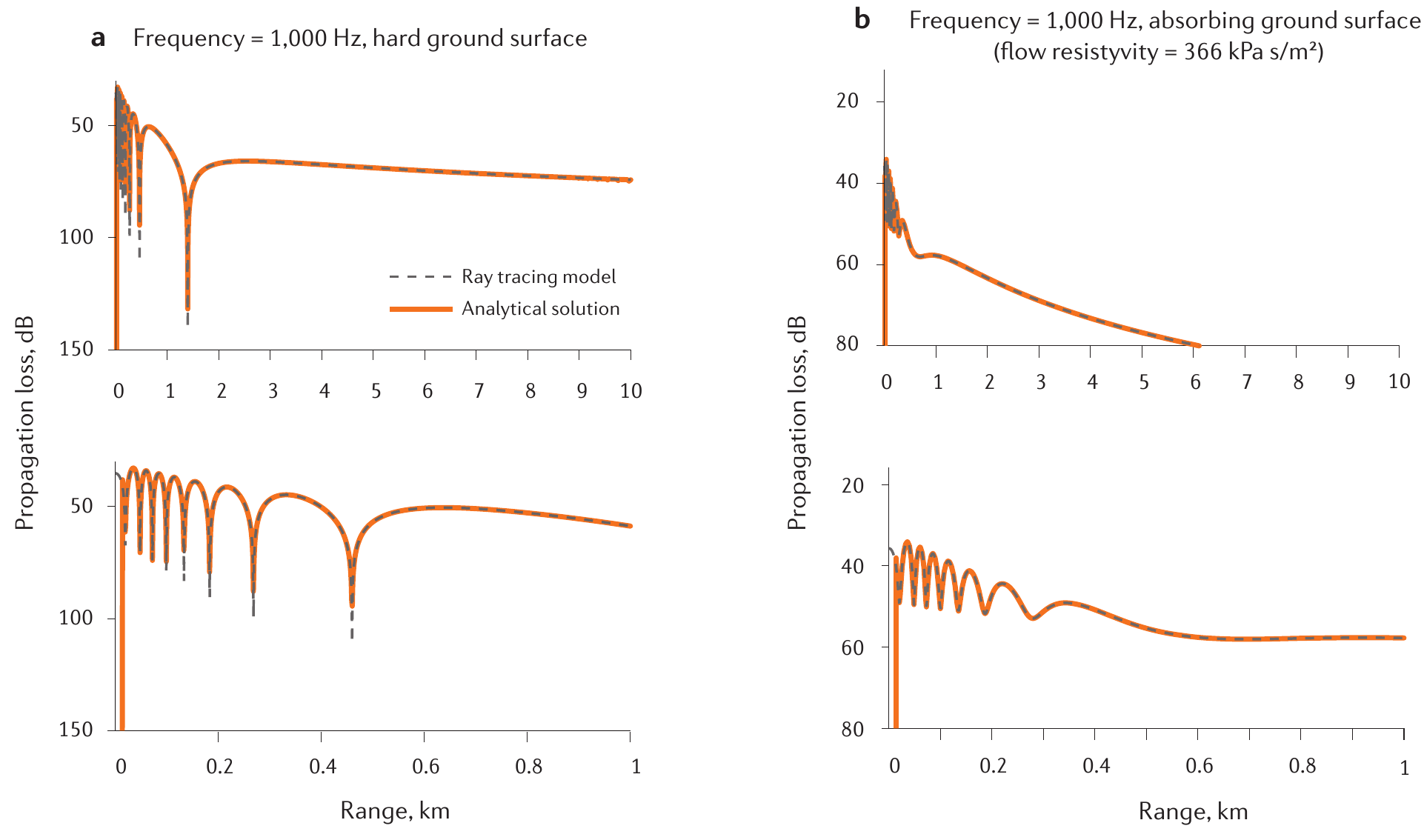}
\end{center}
\caption{{Validations of ray tracing results with analytical solution.}}
\label{fig:FIG2}
\end{figure}

The parameters of the ray tracing model such as the number of rays, launching angle range and beam type are shown in \autoref{tab:table5}. These parameters were determined as suggested in \cite{hansen2019investigation} and our convergence analysis (See Supplementary Fig. S2-S5). We used several approaches to validate the Bellhop model such as comparing with analytical solutions, results in atmospheric sound propagation textbooks \cite{salomons2001computational, ostashev2015acoustics}, and the benchmark case results \cite{attenborough1995benchmark}. In general, the results using Bellhop were comparable to analytical and previously published results, indicating that Bellhop is reliable for atmospheric sound propagation applications. \autoref{fig:FIG2} shows the validation of Bellhop using analytical solutions. The analytical case is a point source at 80 m and a receiver at 1.5 m above ground level. The prediction range is up to 10 km. The source emits a tone at frequencies of 10, 100 and 1000 Hz. The sound speed is constant at 343 m/s. We validated the model for two cases with different ground surface conditions. For case 1, a perfectly hard ground surface was used, in which the plane wave reflection coefficient was set to $R = 1$. For case 2, an absorbing ground surface was used with ground surface parameters similar to values in Table I in \cite{attenborough1995benchmark} (i.e., flow resistivity of 366 $kPa s m^{-2}$, porosity of 0.27, pore shape factor of 0.25 and grain shape factor of 0.5).
Other validation results are provided in detail in Supplementary Fig. S6-13.

\begin{table}[H]
\centering
\caption{\label{tab:table5}Ray tracing model parameters.}

\begin{tabular}{p{6cm} p{3cm}}
\hline\hline
Parameter & Value (Type)\\
\hline
Number of rays &16001 \\
Angle step & 0.01\\
Launching angle& $[-80^o, 80^o]$\\
Beam shape& Gaussian shape\\
Pressure contribution& Coherent\\
Ray step (ray segment length) & 1.0 m\\
Sound speed interpolation& Linear\\

\hline\hline
\end{tabular}
\end{table}

\subsection{Statistical analysis}
\subsubsection{Sensitivity analysis}

To estimate the relative importance of each input parameter on the output of the ground impedance models, global sensitivity analysis based on the Sobol method was implemented \cite{saltelli2010variance}. The input parameters for the ground impedance models included parameters as shown in Table 2, while the output parameters were the real part of specific ground surface impedance values. The resulting information regarding the relative importance of input parameters was used to reduce simulation complexity.

Global sensitivity analysis is a variance-based method that decomposes the variance of the model outputs into fractions contributed by each input parameter and their interactions. Standardised to a total variance of unity provides fractional variance contributions as follows:

\begin{equation} \label{eq2}
\begin{split}
&\sum_{i} S_i + \sum_{i} \sum_{j>i} S_{ij} + ... + S_{12...k} = 1,
\end{split}
\end{equation}
where $S_i$ measures the main effect of the $ i$-th parameter on the output model and $S_{ij}$ denotes the high-order interaction indices between input parameters. The total effect of the $i$-th parameter, $S_{Ti}$, can be simply considered as the main effect $S_i$ plus all interaction terms including the $i$-th parameter. An input parameter with a higher value of $S_{Ti}$ is considered to have higher importance.

\subsubsection{Partition uncertainty}

The uncertainty of noise prediction in the analysis undertaken here arises from uncertainties in the impedance models and atmospheric vertical wind speed profile models. To decompose the total uncertainty into contributing sources, analysis of variance (ANOVA) was used, as proposed by \citet{yip2011simple}. The total uncertainty $T(r)$ as a function of range $r$, is simply the variance of the predicted noise levels, defined as:

\begin{equation} \label{eq3}
\begin{split}
T(r) = ImpMl(r) + WindMl(r) + I(r),
\end{split}
\end{equation}
where $ImpMl(r)$, $WindMl(r)$ and $I(r)$ are the uncertainties due to impedance models, atmospheric vertical wind speed profile models and their interaction, respectively. The proportion of uncertainty contribution from each source is estimated and normalised by the total uncertainty $T(r)$. 

\subsubsection{Statistical tests}

All statistical analysis and data visualisation were implemented in R programming language (\href{http://www.r-project.org/}{http://www.r-project.org/}). Sensitivity analysis based on the Sobol method was implemented using packages `multisensi' and `sensivivity'. Partition uncertainty was implemented by the authors as explained in section 2.6. Linear regression was performed using the R base. The significance threshold used was $\alpha$ = 0.05.

\subsection{Data and code availability}
\label{sub:code_avail}
Simulations were run in parallel on the Flinders DeepThought HPC computer using a single node with 128 logical processors with 500 GB  RAM. Our wrapper package, called FreeRay, written in Julia code to implement BELLHOP, is provided at \href{https://github.com/ducphucnguyen/FreeRay.jl}{https://github.com/ducphucnguyen/FreeRay.jl}

\section{Results}
\subsection{Sensitivity analysis and model discrepancy}

The flow resistivity and porosity of the ground were the most influential parameters that significantly affected the model output (i.e., characteristic impedance, then affecting the reflection coefficient). An example of sensitivity analysis for the four parameter ground impedance model (Imp9) \cite{attenborough1985acoustical} is shown in \autoref{fig:FIG3a}\textbf{a}. Similar results were obtained for other models and are provided in Supplementary Fig. S14-S15. Flow resistivity was the most influential parameter on uncertainty at frequencies less than 800 Hz, while porosity was more influential at higher frequencies, in agreement with \citet{attenborough2002review}. The pore shape factor was less influential, and tortuousity only contributed as a minor factor at high frequencies. Variation between ground impedance models is shown in \autoref{fig:FIG3a}\textbf{b}.  A large discrepancy in the reflection coefficient was observed, especially at higher grazing angles, indicating a possible source of uncertainty. The largest difference was between the variable porosity model (Imp2) and Attenborough's four parameter model (Imp9).
 
 \begin{figure}[H]
\begin{center}
\includegraphics[width=14cm]{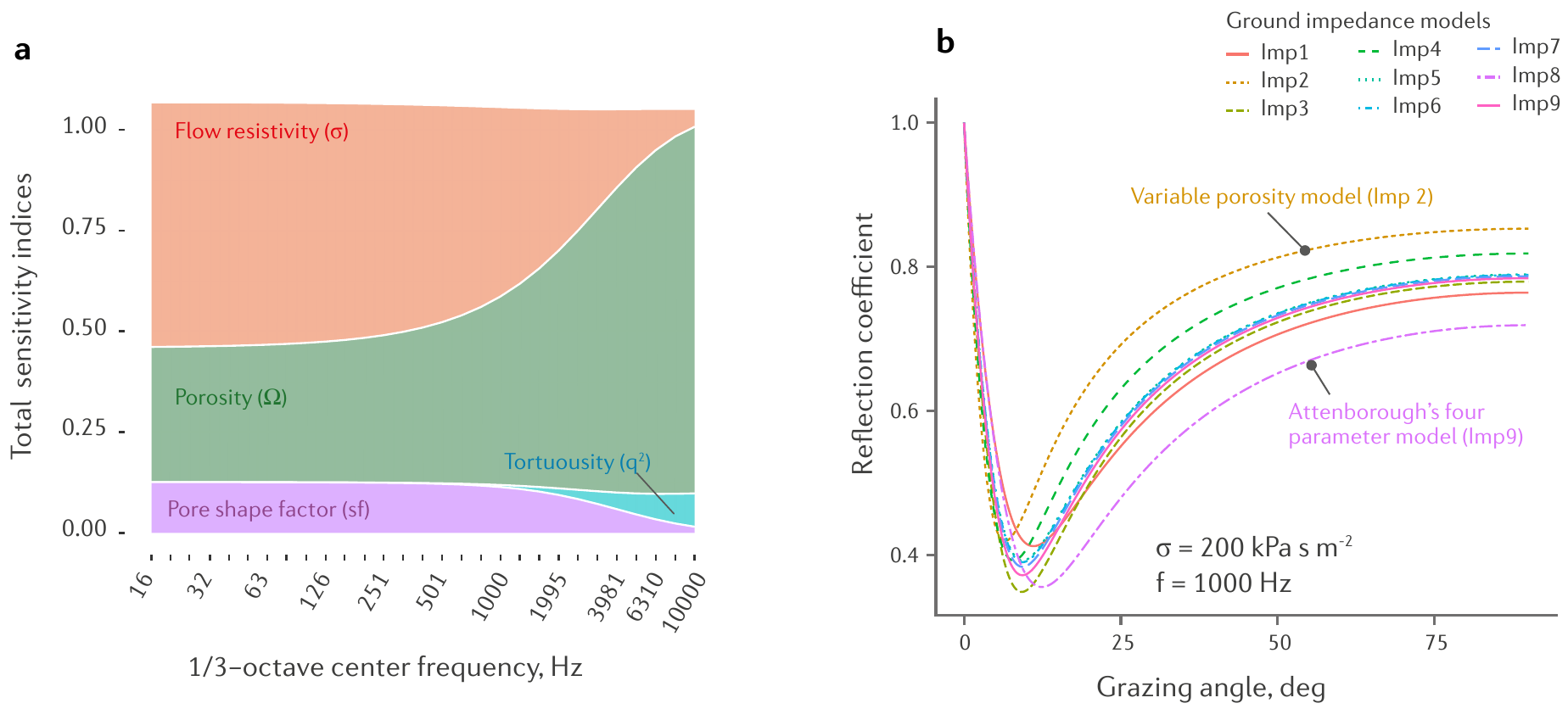}
\end{center}
\caption{{Ground impedance model. \textbf{a}, Sensitivity analysis for Attenborough's four parameter model \cite{attenborough1985acoustical}. \textbf{b}, Ground reflection coefficients for all impedance models using the same set of input parameters.}}
\label{fig:FIG3a}
\end{figure}

\begin{figure}[H]
\begin{center}
\includegraphics[width=14cm]{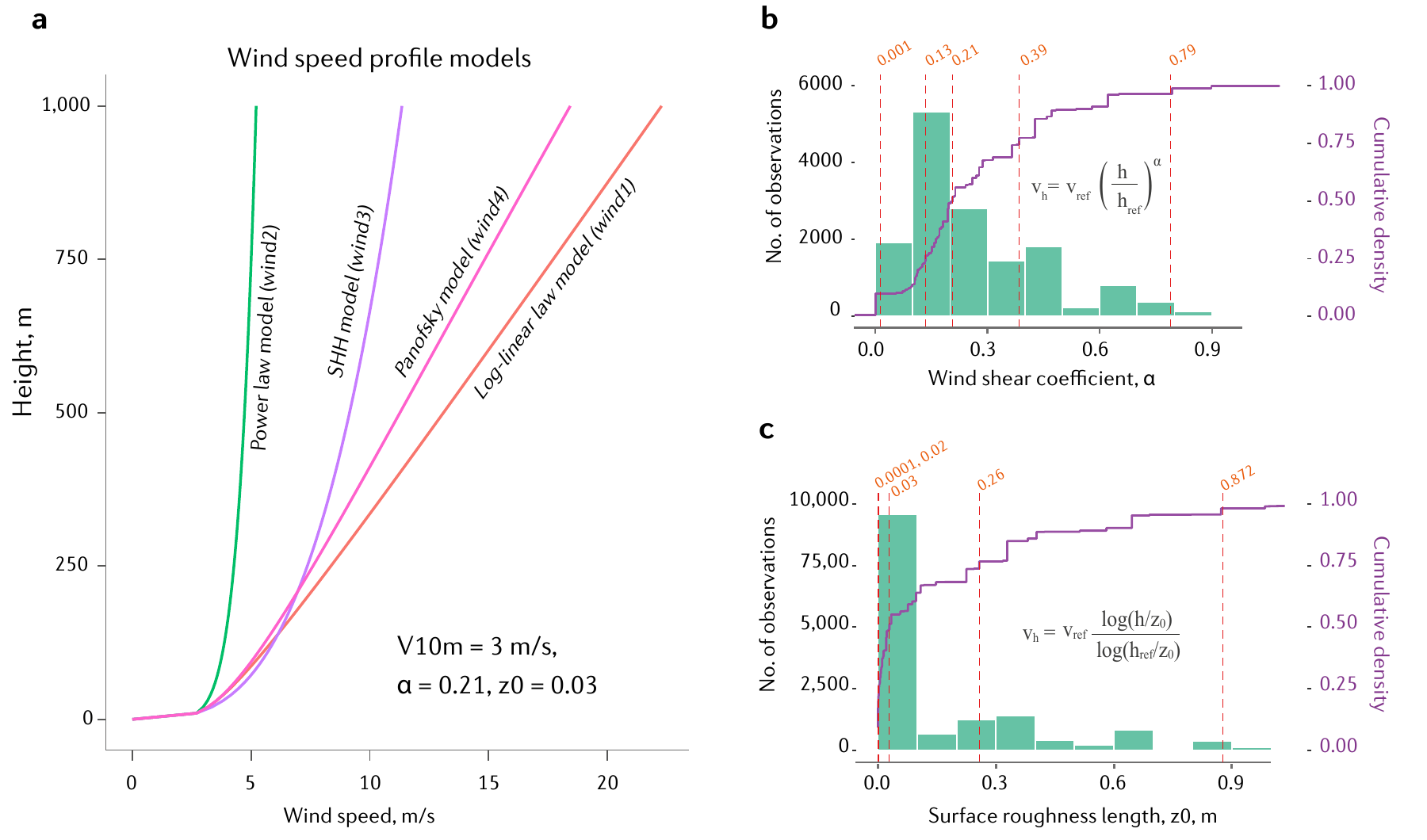}
\end{center}
\caption{{Wind speed profile model. \textbf{a, b}, The distributions of wind shear coefficient, $\alpha$, and surface roughness length, $z_0$, were constructed by substituting the characteristic values obtained from our one year data set into the corresponding equations as shown in the figure. \textbf{c}, Wind speed profiles estimated from wind extrapolation models using the same set of input parameters. }}
\label{fig:FIG3b}
\end{figure}

 Similarly, a large discrepancy between wind profiles was observed when using different wind extrapolation models (\autoref{fig:FIG3b}\textbf{a}). Although both the power law model and log-linear law model are most often used by the wind industry, the difference between these models was the largest.
Wind shear coefficient and surface roughness length were calculated from our experimental data \cite{nguyen2021long} as shown in \autoref{fig:FIG3b}\textbf{b,c}. In our simulation (see \autoref{fig:FIG1}), we used the values at 2.5$^{\rm{th}}$, 25$^{\rm{th}}$, 50$^{\rm{th}}$ and 75$^{\rm{th}}$ 97.5$^{\rm{th}}$ percentiles.

\subsection{Uncertainty in predicting noise levels}

\begin{figure}[H]
\begin{center}
\includegraphics[width=14cm]{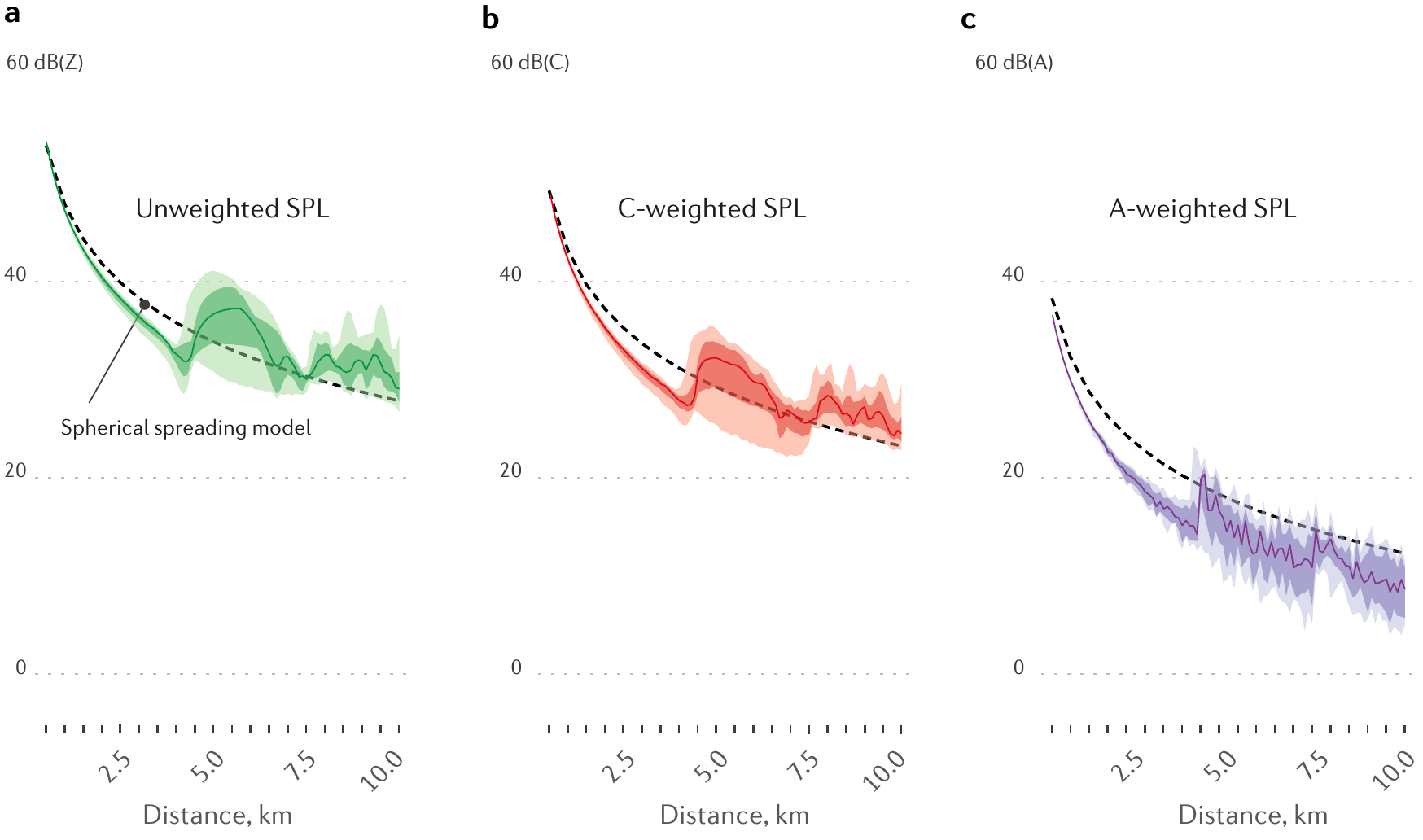}
\end{center}
\caption{{Uncertainty due to input models. \textbf{a, b and c}, A-, C- and unweighted SPLs and their variations due to model uncertainty. The area with darker colour is interquartile range, while lighter colour is between $2.5^{th}$ and $97.5^{th}$ range. The line in the middle of the area show median line. All input parameters for the propagation model were fixed. Flow resistivity $\sigma= 200$ kPa s m$^{-2}$, wind shear coefficient $\alpha = 0.21$ and surface roughness length $z_0 = 0.03$, corresponding to median levels.}}
\label{fig:FIG4a}
\end{figure}

Predicted noise level uncertainty increased with distance for all three metrics as shown in  \autoref{fig:FIG4a}. Predicted noise level variations were smaller at distances less than 1 km, but the variation was substantial at distances greater than 4 km. The noise level at long-range distances was dominated by influences associated with refracted rays \cite{ostashev2015acoustics}. These refracted rays occurred due to the positive wind profiles (increase in wind velocity as a function of height).  Thus, the larger variation between the atmospheric vertical wind speed profile model outputs could be attributed to the interaction between refracted rays, which results in higher variations of predicted noise levels. For reference, the predicted noise levels using the spherical spreading law were also calculated as shown in dashed curves in \autoref{fig:FIG4a}. This model depends on distance only, and thus these curves were considered as the maximum predicted noise levels in an unbounded homogeneous medium \cite{jensen2011computational}.

\begin{figure}[H]
\begin{center}
\includegraphics[width=14cm]{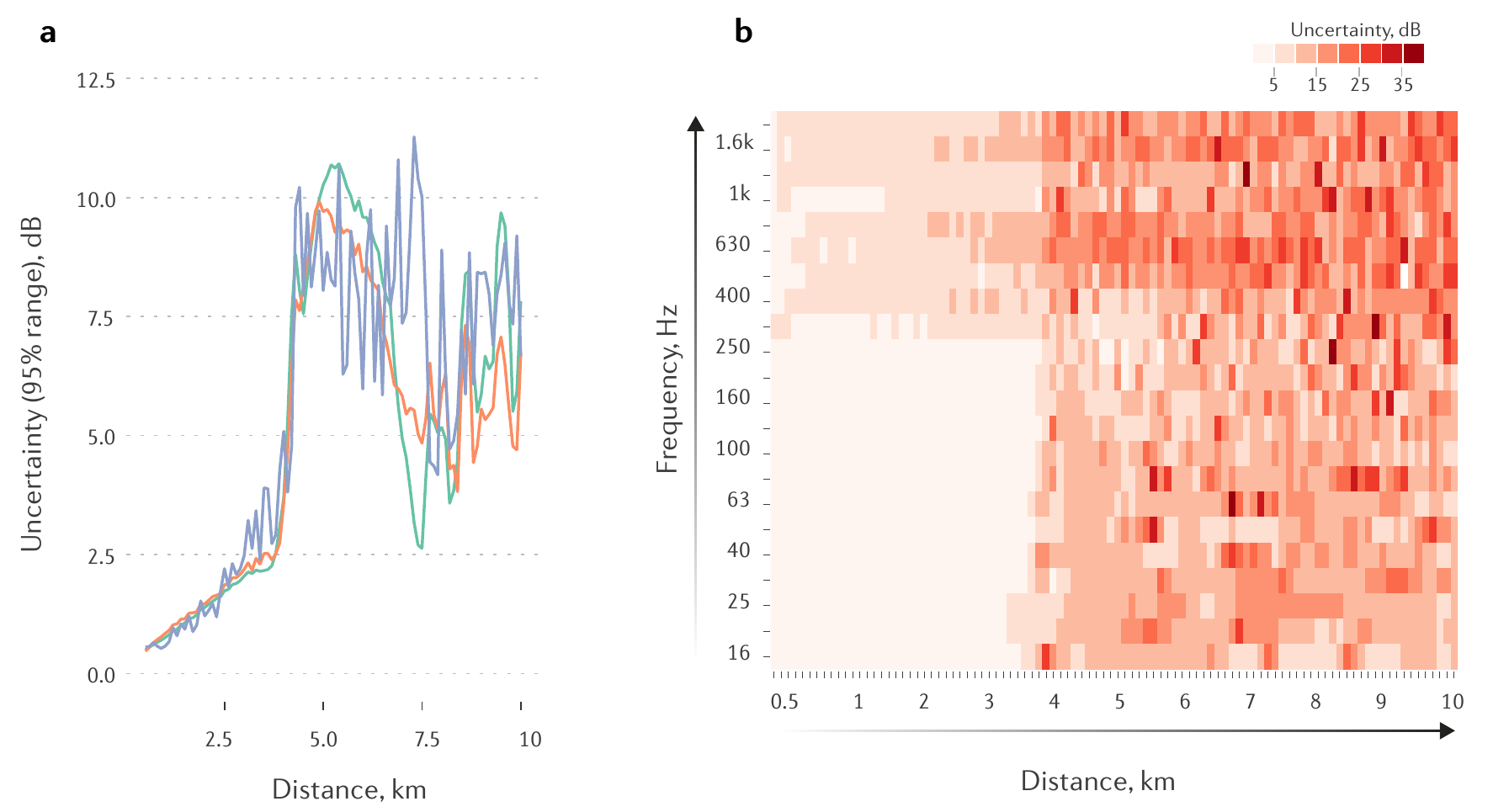}
\end{center}
\caption{{Uncertainty due to input models. \textbf{a}, Relationship between uncertainty and distance for A-, C- and unweighted SPLs.  The uncertainty is quantified as the difference between the $2.5\%$ and $97.5\%$ of the predicted noise level variation (95\% confidence interval). All input parameters for the propagation model were fixed as specified in \autoref{fig:FIG4a}.}}
\label{fig:FIG4b}
\end{figure}

To further quantify uncertainty, the variation (95\% range) of the predicted noise levels are shown in \autoref{fig:FIG4b}\textbf{a}. All three metrics showed similar patterns. For example, the variation increased linearly between 0.5 and 3.5 km at a rate of 0.8 dB/km. Between 4 and 10 km, the variation was generally between around 3.5 and 10 dB, although local peaks and troughs were observed outside of this range. The uncertainty in predicting A-weighted SPLs could be above 10 dBA at particular locations, indicating strong constructive and destructive interference between  direct and reflected waves \cite{ostashev2015acoustics}. These detailed patterns would not be well predicted using typical engineering models to predict wind farm noise \cite{ISO9613}.

The relationships between uncertainty, distance and frequency are shown in \autoref{fig:FIG4b}\textbf{b}. Higher uncertainty was observed at higher frequencies and longer-range receivers. This indicates that uncertainty quantification is highly dependent on the frequency content of noise sources. Specifically, if the noise source spectrum is dominated by low frequencies, the predicted noise level is likely to have a lower associated uncertainty compared to a noise source with a dominant high frequency content.

\subsection{Partition uncertainty}

\begin{figure}[H]
\begin{center}
\includegraphics[width=14cm]{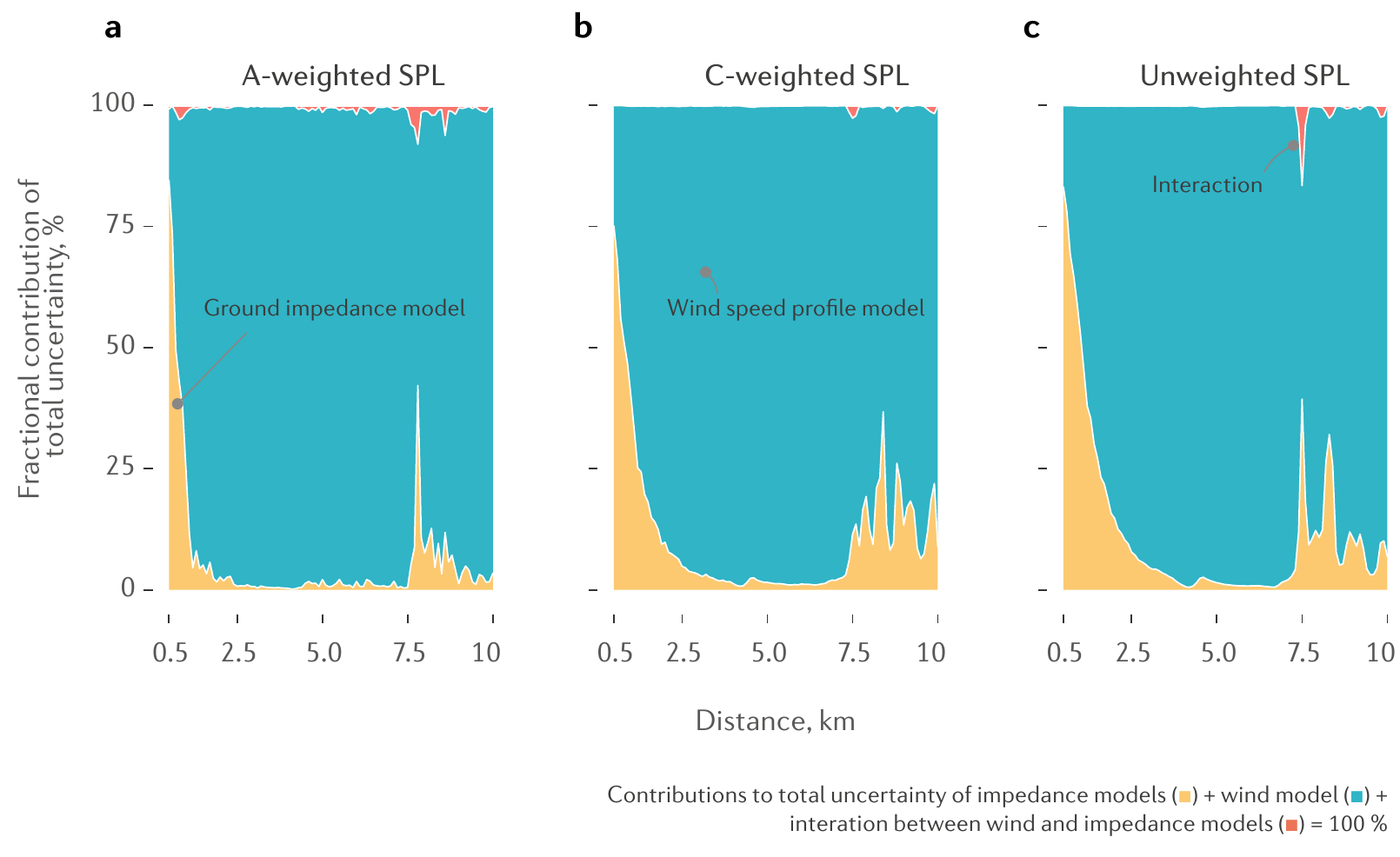}
\end{center}
\caption{{Partition uncertainty analysis. Each vertical slice of the plot represents the contribution of each uncertainty source to the total uncertainty. For example, in figure (a), at a distance of 0.5 km, the uncertainty in predicted noise level comprises 80\% and 20\% contributions from the discrepancies associated with the ground impedance models and wind speed profile models, respectively, while the contribution associated with interaction between these models is small.}}
\label{fig:FIG5a}
\end{figure}

To identify the contribution of each uncertainty source to the variation of predicted noise levels, partition uncertainty analysis was implemented (See Methods section 2.5.2), and the results are shown in \autoref{fig:FIG5a}. The total variation was decomposed into each uncertainty source such as atmospheric vertical wind speed profile models, ground impedance models and their interaction. The contribution of impedance models to the total uncertainty was significantly reduced with increasing distance. Impedance models contributed $>$75\% of total uncertainty of predicted A-weighted SPLs up to 0.5 km, which reduced to $<$ 25\% at 1 km (\autoref{fig:FIG5a}\textbf{a}). Similar trends were also  observed for predicted C- and unweighted SPLs (\autoref{fig:FIG5a}\textbf{b, c}), but the reduction in uncertainty occurred  at a lower rate, indicating that impedance models were likely to have a greater effect on the low frequency content. In contrast, atmospheric vertical wind speed profile models were the main source of uncertainty at long-range distances. At distances greater than 2.5 km, this source contributed to $>$ 95\%, 90\% and 85\% of the variation for predicted A-, C- and unweighted SPLs, respectively. The interaction between these two uncertainty sources was small as shown in red regions in \autoref{fig:FIG5a}. 

\subsection{Interaction with input value}

\begin{figure}[H]
\begin{center}
\includegraphics[width=14cm]{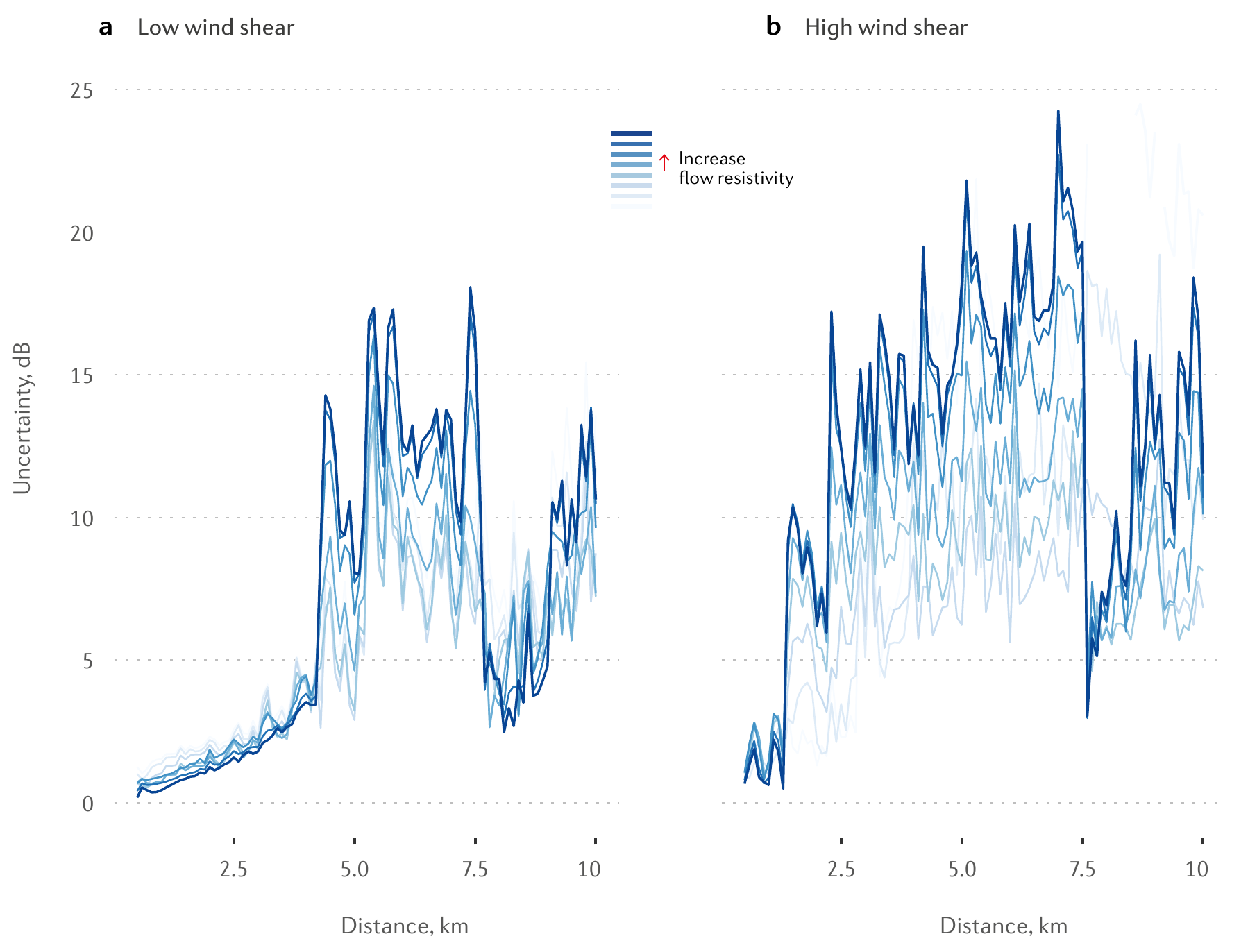}
\end{center}
\caption{{Interaction between input values such as flow resistivity and wind shear coefficient (or roughness length). \textbf{a}, Low wind shear coefficient was combined with eight levels of flow resistivity. Each combination was included as an input to the propagation model to quantify the uncertainty. \textbf{b},  high wind shear states were combined with eight levels of flow resistivity. }}
\label{fig:FIG5b}
\end{figure}

To further investigate if changing the flow resistivity and wind shear coefficient (or ground roughness length) can affect uncertainty in predicting A-weighted SPL, eight levels of flow resistivity were combined with five levels of refractive state. Each combination went through the uncertainty propagation pipeline (\autoref{fig:FIG1}) in order to quantify the level of uncertainty, as shown in \autoref{fig:FIG5b}. Uncertainty increased with increasing flow resistivity, $\sigma$, wind shear coefficient, $\alpha$, and distance (linear regression, all $P$-value $<$ 0.001, $R^2$ = 0.45), fitted linear equations as follows:

\begin{equation} \label{eq4}
\begin{split}
\textrm{Uncertainty~(dB)} = -1.6\times 10^{-2} +  8.4\times10^{-6} \sigma + 6.3 \alpha + 1.0 \times \textrm{distance}
\end{split}
\end{equation}

Changing the flow resistivity did not substantially impact uncertainty. For example, uncertainty increased by 1.7 dB when flow resistivity increased from 12.5 to 200,000 kPa s $m^{-2}$. In contrast, uncertainty increased by approximately 5 dB when the wind shear coefficient increased from 0.001 to 0.79. The uncertainty was also much more sensitive to changes in distance, increasing by 9.5 dB from 0.5 to 10 km. Eq. (\ref{eq4}) could be used to estimate the uncertainty associated with input models in wind farm noise prediction.

\section{Discussion}

Variations between different impedance and atmospheric vertical wind speed profile models were identified as non-negligible uncertainty sources, especially for long-range noise predictions. We also found that atmospheric vertical wind speed profile model variations had greater effects on the uncertainty in noise level predictions than variations in ground impedance models at long-range distances.

The discrepancy between input models results in large uncertainties in predicting outdoor noise levels. The uncertainty can reach over 10 dBA at distances greater than 4 km. As impedance models differ in the way they use input variables, set parameters and use approximations, large discrepancies between these models in calculating impedance have
been reported \cite{salomons2012computational,komatsu2008improvement,taraldsen2005delany,wilson1997simple,ostashev2015acoustics}. Although recommendations for using a suitable impedance model for outdoor sound propagation applications have been made \cite{attenborough2011outdoor,ostashev2015acoustics,wilson1997simple}, a range of different models are still used for predicting outdoor noise. The single parameter semi-empirical model proposed by \citet{delany1970acoustical} has been used widely to characterise outdoor ground surfaces, although this model can be erroneous for low frequencies in some cases \cite{taraldsen2005delany}. Similarly, logarithmic wind profile models are now considered unsuitable for calculating the wind profile compared to power law models \cite{gualtieri2019comprehensive}, although both models are still used for outdoor noise prediction \cite{Hansen2017,bies2017engineering,plovsing1849nord2000}. Suggesting a suitable impedance model or wind profile model for outdoor noise prediction is beyond the scope of the current study, given that uncertainty of predictions is not necessarily a reflection of model accuracy. However, these findings highlight that discrepancies between models can be a large source of uncertainty. 

Uncertainty associated with model discrepancy at long-range locations is mainly due to atmospheric vertical wind speed profile models. This might be due to the contribution of sound refraction at long range locations \cite{Hansen2017,ostashev2015acoustics}. In contrast, the effect of ground impedance models on the uncertainty of long-range noise prediction is small. This is likely because sound waves are significantly attenuated by the ground after reflecting multiple times. Our findings suggest that for long-range noise predictions, more focus should be placed on determining the atmospheric vertical wind speed profile and other meteorological conditions. These parameters should be estimated by measurement or by using more sophisticated models, such as a dedicated micrometeorological code (SUBMESO, \cite{lihoreau2006outdoor}) which simulates wind and temperature fields over moderately complex terrain with high resolution. 

A limitation of the present study is that only uncertainty in the downwind direction was investigated. The simplification of geometry is a further limitation, although this makes the problem more tractable and general. We were also unable to comprehensively consider other parameters known to significantly influence noise predictions, including temperature profiles, atmospheric turbulence and noise source characteristics \cite{ostashev2015acoustics}. We also did not consider variations in ground impedance and wind speed profiles that could occur as noise propagates from the source to the receiver. Another limitation of our study is that the uncertainty associated with the discrepancy between sound propagation models was not quantified. This could be one of the most important sources of uncertainty as a large difference between models has been found in \cite{salomons2001computational,salomons1994downwind,heimann1999coupled,mo2017outdoor}. A comprehensive uncertainty quantification which considers both parametric and model structure uncertainty is needed to comprehensively understand variations in noise predictions. The ray tracing model is efficient and more suitable for uncertainty quantification where calculation speed is important \cite{jensen2011computational}. However, it contains potentially problematic high frequency approximations and caustics \cite{ostashev2015acoustics}. Although the problems with caustics in ray models can be reduced by implementing the Gaussian beam tracing approach, as in Bellhop (\cite{jensen2011computational} , page 180), the problem is not completely addressed and thus wave-based methods such as the PE model should be considered in future studies to improve model performance.

\section{Conclusions}

We conclude that the model structure uncertainty associated with multi-input models is significant. Our study is the first attempt to quantify this systematic source of uncertainty that is typically overlooked in quantifying outdoor noise prediction uncertainty. The level of uncertainty increases with distance between the source and receiver and can exceed 10 dBA at distances greater than 3.5 km for cases with high wind shear and high ground flow resistivity.  We also find that uncertainty increases with increasing wind shear coefficient and ground flow resistivity. We have provided a quantitative contribution of each uncertainty source to the variation of predicted noise levels. We found that the uncertainty associated with ground impedance models was small at distances greater than 1 km, while most of the variation in predicted SPL was attributed to vertical wind speed profile models, suggesting that good knowledge of the refractive state of the atmosphere is mandatory when predicting WFN noise at distances greater than 1 km. Our approach presented here could be extended to quantify other systematic sources of uncertainty such as that arising from the use of different sound propagation models. Combining model structure uncertainty and parametric uncertainty arising from imperfect models of complex systems, outdoor sound prediction practice could be advanced to a more unified and probabilistic approach that has been successfully applied in other fields such as weather forecasting.

\section*{Acknowledgements}
The authors gratefully acknowledge financial support from the Australian Research Council, projects DP120102185 and DE180100022 and the National Health and Medical Research Council, Project 1113571. The author DPN was supported by a Flinders University Research Scholarship (FURS) for this work.

\bibliographystyle{elsarticle-num-names} 
\bibliography{WTN_all_V1.bib}

\begin{thebibliography}{51}
\expandafter\ifx\csname natexlab\endcsname\relax\def\natexlab#1{#1}\fi
\providecommand{\url}[1]{\texttt{#1}}
\providecommand{\href}[2]{#2}
\providecommand{\path}[1]{#1}
\providecommand{\DOIprefix}{doi:}
\providecommand{\ArXivprefix}{arXiv:}
\providecommand{\URLprefix}{URL: }
\providecommand{\Pubmedprefix}{pmid:}
\providecommand{\doi}[1]{\href{http://dx.doi.org/#1}{\path{#1}}}
\providecommand{\Pubmed}[1]{\href{pmid:#1}{\path{#1}}}
\providecommand{\bibinfo}[2]{#2}
\ifx\xfnm\relax \def\xfnm[#1]{\unskip,\space#1}\fi
\bibitem[{Hansen et~al.(2018)Hansen, Zajamsek, and Hansen}]{hansen2018wind}
\bibinfo{author}{K.~L. Hansen}, \bibinfo{author}{B.~Zajamsek},
  \bibinfo{author}{C.~H. Hansen},
\newblock \bibinfo{title}{Wind farm noise uncertainty: Prediction, measurement
  and compliance assessment},
\newblock \bibinfo{journal}{Acoustics Australia} \bibinfo{volume}{46}
  (\bibinfo{year}{2018}) \bibinfo{pages}{59--67}.
\bibitem[{Salomons(2012)}]{salomons2012computational}
\bibinfo{author}{E.~M. Salomons}, \bibinfo{title}{Computational atmospheric
  acoustics}, \bibinfo{publisher}{Springer Science \& Business Media},
  \bibinfo{year}{2012}.
\bibitem[{Ostashev and Wilson(2015)}]{ostashev2015acoustics}
\bibinfo{author}{V.~E. Ostashev}, \bibinfo{author}{D.~K. Wilson},
  \bibinfo{title}{Acoustics in moving inhomogeneous media},
  \bibinfo{publisher}{CRC Press}, \bibinfo{year}{2015}.
\bibitem[{Jensen et~al.(2011)Jensen, Kuperman, Porter, and
  Schmidt}]{jensen2011computational}
\bibinfo{author}{F.~B. Jensen}, \bibinfo{author}{W.~A. Kuperman},
  \bibinfo{author}{M.~B. Porter}, \bibinfo{author}{H.~Schmidt},
  \bibinfo{title}{Computational ocean acoustics}, \bibinfo{publisher}{Springer
  Science \& Business Media}, \bibinfo{year}{2011}.
\bibitem[{Wilson et~al.(2007)Wilson, Andreas, Weatherly, Pettit, Patton, and
  Sullivan}]{wilson2007characterization}
\bibinfo{author}{D.~K. Wilson}, \bibinfo{author}{E.~L. Andreas},
  \bibinfo{author}{J.~W. Weatherly}, \bibinfo{author}{C.~L. Pettit},
  \bibinfo{author}{E.~G. Patton}, \bibinfo{author}{P.~P. Sullivan},
\newblock \bibinfo{title}{Characterization of uncertainty in outdoor sound
  propagation predictions},
\newblock \bibinfo{journal}{The Journal of the Acoustical Society of America}
  \bibinfo{volume}{121} (\bibinfo{year}{2007}) \bibinfo{pages}{EL177--EL183}.
\bibitem[{Van~Renterghem and Botteldooren(2018)}]{van2018variability}
\bibinfo{author}{T.~Van~Renterghem}, \bibinfo{author}{D.~Botteldooren},
\newblock \bibinfo{title}{Variability due to short-distance favorable sound
  propagation and its consequences for immission assessment},
\newblock \bibinfo{journal}{The Journal of the Acoustical Society of America}
  \bibinfo{volume}{143} (\bibinfo{year}{2018}) \bibinfo{pages}{3406--3417}.
\bibitem[{Wilson et~al.(2014)Wilson, Pettit, Ostashev, and
  Vecherin}]{wilson2014description}
\bibinfo{author}{D.~K. Wilson}, \bibinfo{author}{C.~L. Pettit},
  \bibinfo{author}{V.~E. Ostashev}, \bibinfo{author}{S.~N. Vecherin},
\newblock \bibinfo{title}{Description and quantification of uncertainty in
  outdoor sound propagation calculations},
\newblock \bibinfo{journal}{The Journal of the Acoustical Society of America}
  \bibinfo{volume}{136} (\bibinfo{year}{2014}) \bibinfo{pages}{1013--1028}.
\bibitem[{Attenborough et~al.(1995)Attenborough, Taherzadeh, Bass, Di, Raspet,
  Becker, G{\"u}desen, Chrestman, Daigle, L’Esp{\'e}rance
  et~al.}]{attenborough1995benchmark}
\bibinfo{author}{K.~Attenborough}, \bibinfo{author}{S.~Taherzadeh},
  \bibinfo{author}{H.~E. Bass}, \bibinfo{author}{X.~Di},
  \bibinfo{author}{R.~Raspet}, \bibinfo{author}{G.~Becker},
  \bibinfo{author}{A.~G{\"u}desen}, \bibinfo{author}{A.~Chrestman},
  \bibinfo{author}{G.~A. Daigle}, \bibinfo{author}{A.~L’Esp{\'e}rance},
  et~al.,
\newblock \bibinfo{title}{Benchmark cases for outdoor sound propagation
  models},
\newblock \bibinfo{journal}{The Journal of the Acoustical Society of America}
  \bibinfo{volume}{97} (\bibinfo{year}{1995}) \bibinfo{pages}{173--191}.
\bibitem[{Kayser et~al.(2020)Kayser, Cott{\'e}, Ecoti{\`e}re, and
  Gauvreau}]{kayser2020environmental}
\bibinfo{author}{B.~Kayser}, \bibinfo{author}{B.~Cott{\'e}},
  \bibinfo{author}{D.~Ecoti{\`e}re}, \bibinfo{author}{B.~Gauvreau},
\newblock \bibinfo{title}{Environmental parameters sensitivity analysis for the
  modeling of wind turbine noise in downwind conditions},
\newblock \bibinfo{journal}{The Journal of the Acoustical Society of America}
  \bibinfo{volume}{148} (\bibinfo{year}{2020}) \bibinfo{pages}{3623--3632}.
\bibitem[{Parry et~al.(2020)Parry, Horoshenkov, and
  Williams}]{parry2020investigating}
\bibinfo{author}{J.~A. Parry}, \bibinfo{author}{K.~V. Horoshenkov},
  \bibinfo{author}{D.~P. Williams},
\newblock \bibinfo{title}{Investigating uncertain geometries effect on sound
  propagation in a homogeneous and non-moving atmosphere over an impedance
  ground},
\newblock \bibinfo{journal}{Applied Acoustics} \bibinfo{volume}{160}
  (\bibinfo{year}{2020}) \bibinfo{pages}{107122}.
\bibitem[{H{\"o}rmeyer et~al.(2019)H{\"o}rmeyer, H{\"u}bler, Bohne, and
  Rolfes}]{hormeyer2019prediction}
\bibinfo{author}{J.~H{\"o}rmeyer}, \bibinfo{author}{C.~J. H{\"u}bler},
  \bibinfo{author}{T.~Bohne}, \bibinfo{author}{R.~Rolfes},
  \bibinfo{title}{Prediction of atmospheric sound propagation subject to
  parameter variability of atmospheric turbulence},
  \bibinfo{publisher}{Universit{\"a}tsbibliothek der RWTH Aachen},
  \bibinfo{year}{2019}.
\bibitem[{Leroy et~al.(2010)Leroy, Gauvreau, Junker, De~Rocquigny, Berengier
  et~al.}]{leroy2010uncertainty}
\bibinfo{author}{O.~Leroy}, \bibinfo{author}{B.~Gauvreau},
  \bibinfo{author}{F.~Junker}, \bibinfo{author}{E.~De~Rocquigny},
  \bibinfo{author}{M.~Berengier}, et~al.,
\newblock \bibinfo{title}{Uncertainty assessment for outdoor sound
  propagation},
\newblock in: \bibinfo{booktitle}{International Congress on Acoustics (ICA)},
  \bibinfo{year}{2010}.
\bibitem[{Pettit and Wilson(2007)}]{pettit2007proper}
\bibinfo{author}{C.~L. Pettit}, \bibinfo{author}{D.~K. Wilson},
\newblock \bibinfo{title}{Proper orthogonal decomposition and cluster weighted
  modeling for sensitivity analysis of sound propagation in the atmospheric
  surface layer},
\newblock \bibinfo{journal}{The Journal of the Acoustical Society of America}
  \bibinfo{volume}{122} (\bibinfo{year}{2007}) \bibinfo{pages}{1374--1390}.
\bibitem[{Kayser et~al.(2019)Kayser, Gauvreau, and
  Ecoti{\`e}re}]{kayser2019sensitivity}
\bibinfo{author}{B.~Kayser}, \bibinfo{author}{B.~Gauvreau},
  \bibinfo{author}{D.~Ecoti{\`e}re},
\newblock \bibinfo{title}{Sensitivity analysis of a parabolic equation model to
  ground impedance and surface roughness for wind turbine noise},
\newblock \bibinfo{journal}{The Journal of the Acoustical Society of America}
  \bibinfo{volume}{146} (\bibinfo{year}{2019}) \bibinfo{pages}{3222--3231}.
\bibitem[{Attenborough et~al.(2011)Attenborough, Bashir, and
  Taherzadeh}]{attenborough2011outdoor}
\bibinfo{author}{K.~Attenborough}, \bibinfo{author}{I.~Bashir},
  \bibinfo{author}{S.~Taherzadeh},
\newblock \bibinfo{title}{Outdoor ground impedance models},
\newblock \bibinfo{journal}{The Journal of the Acoustical Society of America}
  \bibinfo{volume}{129} (\bibinfo{year}{2011}) \bibinfo{pages}{2806--2819}.
\bibitem[{Gualtieri(2019)}]{gualtieri2019comprehensive}
\bibinfo{author}{G.~Gualtieri},
\newblock \bibinfo{title}{A comprehensive review on wind resource extrapolation
  models applied in wind energy},
\newblock \bibinfo{journal}{Renewable and Sustainable Energy Reviews}
  \bibinfo{volume}{102} (\bibinfo{year}{2019}) \bibinfo{pages}{215--233}.
\bibitem[{Asseng et~al.(2013)Asseng, Ewert, Rosenzweig, Jones, Hatfield, Ruane,
  Boote, Thorburn, R{\"o}tter, Cammarano et~al.}]{asseng2013uncertainty}
\bibinfo{author}{S.~Asseng}, \bibinfo{author}{F.~Ewert},
  \bibinfo{author}{C.~Rosenzweig}, \bibinfo{author}{J.~W. Jones},
  \bibinfo{author}{J.~L. Hatfield}, \bibinfo{author}{A.~C. Ruane},
  \bibinfo{author}{K.~J. Boote}, \bibinfo{author}{P.~J. Thorburn},
  \bibinfo{author}{R.~P. R{\"o}tter}, \bibinfo{author}{D.~Cammarano}, et~al.,
\newblock \bibinfo{title}{Uncertainty in simulating wheat yields under climate
  change},
\newblock \bibinfo{journal}{Nature climate change} \bibinfo{volume}{3}
  (\bibinfo{year}{2013}) \bibinfo{pages}{827--832}.
\bibitem[{Tebaldi et~al.(2005)Tebaldi, Smith, Nychka, and
  Mearns}]{tebaldi2005quantifying}
\bibinfo{author}{C.~Tebaldi}, \bibinfo{author}{R.~L. Smith},
  \bibinfo{author}{D.~Nychka}, \bibinfo{author}{L.~O. Mearns},
\newblock \bibinfo{title}{Quantifying uncertainty in projections of regional
  climate change: A bayesian approach to the analysis of multimodel ensembles},
\newblock \bibinfo{journal}{Journal of Climate} \bibinfo{volume}{18}
  (\bibinfo{year}{2005}) \bibinfo{pages}{1524--1540}.
\bibitem[{Porter(2011)}]{porter2011bellhop}
\bibinfo{author}{M.~B. Porter},
\newblock \bibinfo{title}{The bellhop manual and user’s guide: Preliminary
  draft}  (\bibinfo{year}{2011}).
\bibitem[{Hansen et~al.(2017)Hansen, Doolan, and Hansen}]{Hansen2017}
\bibinfo{author}{C.~H. Hansen}, \bibinfo{author}{C.~J. Doolan},
  \bibinfo{author}{K.~L. Hansen}, \bibinfo{title}{Wind Farm Noise: Measurement,
  Assessment and Control}, \bibinfo{edition}{1} ed., \bibinfo{publisher}{John
  Wiley \& Sons Ltd}, \bibinfo{year}{2017}.
\bibitem[{Prospathopoulos and Voutsinas(2007)}]{prospathopoulos2007application}
\bibinfo{author}{J.~M. Prospathopoulos}, \bibinfo{author}{S.~G. Voutsinas},
\newblock \bibinfo{title}{Application of a ray theory model to the prediction
  of noise emissions from isolated wind turbines and wind parks},
\newblock \bibinfo{journal}{Wind Energy: An International Journal for Progress
  and Applications in Wind Power Conversion Technology} \bibinfo{volume}{10}
  (\bibinfo{year}{2007}) \bibinfo{pages}{103--119}.
\bibitem[{Bies et~al.(2017)Bies, Hansen, and Howard}]{bies2017engineering}
\bibinfo{author}{D.~A. Bies}, \bibinfo{author}{C.~Hansen},
  \bibinfo{author}{C.~Howard}, \bibinfo{title}{Engineering noise control},
  \bibinfo{publisher}{CRC press}, \bibinfo{year}{2017}.
\bibitem[{Leonard(2006)}]{Andrew2006}
\bibinfo{author}{A.~Leonard},
\newblock \bibinfo{title}{Noise impact assessment report capital wind farm},
\newblock \bibinfo{journal}{Department of Planning, Industry and Environment,
  NSW Government, Australia}  (\bibinfo{year}{2006}).
\bibitem[{Horoshenkov et~al.(2019)Horoshenkov, Hurrell, and
  Groby}]{horoshenkov2019three}
\bibinfo{author}{K.~V. Horoshenkov}, \bibinfo{author}{A.~Hurrell},
  \bibinfo{author}{J.-P. Groby},
\newblock \bibinfo{title}{A three-parameter analytical model for the acoustical
  properties of porous media},
\newblock \bibinfo{journal}{The Journal of the Acoustical Society of America}
  \bibinfo{volume}{145} (\bibinfo{year}{2019}) \bibinfo{pages}{2512--2517}.
\bibitem[{Delany and Bazley(1970)}]{delany1970acoustical}
\bibinfo{author}{M.~Delany}, \bibinfo{author}{E.~Bazley},
\newblock \bibinfo{title}{Acoustical properties of fibrous absorbent
  materials},
\newblock \bibinfo{journal}{Applied acoustics} \bibinfo{volume}{3}
  (\bibinfo{year}{1970}) \bibinfo{pages}{105--116}.
\bibitem[{Attenborough(2015)}]{attenborough2015outdoor}
\bibinfo{author}{K.~Attenborough},
\newblock \bibinfo{title}{Outdoor ground impedance models},
\newblock \bibinfo{organization}{Euronoise}, \bibinfo{year}{2015}.
\bibitem[{Wilson(1997)}]{wilson1997simple}
\bibinfo{author}{D.~K. Wilson},
\newblock \bibinfo{title}{Simple, relaxational models for the acoustical
  properties of porous media},
\newblock \bibinfo{journal}{Applied Acoustics} \bibinfo{volume}{50}
  (\bibinfo{year}{1997}) \bibinfo{pages}{171--188}.
\bibitem[{Zwikker and Kosten(1949)}]{zwikker1949sound}
\bibinfo{author}{C.~Zwikker}, \bibinfo{author}{C.~W. Kosten},
  \bibinfo{title}{Sound absorbing materials}, \bibinfo{publisher}{Elsevier
  publishing company}, \bibinfo{year}{1949}.
\bibitem[{Taraldsen(2005)}]{taraldsen2005delany}
\bibinfo{author}{G.~Taraldsen},
\newblock \bibinfo{title}{The delany-bazley impedance model and darcy's law},
\newblock \bibinfo{journal}{Acta Acustica united with Acustica}
  \bibinfo{volume}{91} (\bibinfo{year}{2005}) \bibinfo{pages}{41--50}.
\bibitem[{B{\'e}rengier et~al.(1997)B{\'e}rengier, Stinson, Daigle, and
  Hamet}]{berengier1997porous}
\bibinfo{author}{M.~B{\'e}rengier}, \bibinfo{author}{M.~Stinson},
  \bibinfo{author}{G.~Daigle}, \bibinfo{author}{J.~Hamet},
\newblock \bibinfo{title}{Porous road pavements: Acoustical characterization
  and propagation effects},
\newblock \bibinfo{journal}{The Journal of the Acoustical Society of America}
  \bibinfo{volume}{101} (\bibinfo{year}{1997}) \bibinfo{pages}{155--162}.
\bibitem[{Attenborough(1987)}]{attenborough1987acoustic}
\bibinfo{author}{K.~Attenborough},
\newblock \bibinfo{title}{On the acoustic slow wave in air-filled granular
  media},
\newblock \bibinfo{journal}{The Journal of the Acoustical Society of America}
  \bibinfo{volume}{81} (\bibinfo{year}{1987}) \bibinfo{pages}{93--102}.
\bibitem[{Horoshenkov et~al.(2016)Horoshenkov, Groby, and
  Dazel}]{horoshenkov2016asymptotic}
\bibinfo{author}{K.~V. Horoshenkov}, \bibinfo{author}{J.-P. Groby},
  \bibinfo{author}{O.~Dazel},
\newblock \bibinfo{title}{Asymptotic limits of some models for sound
  propagation in porous media and the assignment of the pore characteristic
  lengths},
\newblock \bibinfo{journal}{The journal of the acoustical society of America}
  \bibinfo{volume}{139} (\bibinfo{year}{2016}) \bibinfo{pages}{2463--2474}.
\bibitem[{Monin and Obukhov(1954)}]{monin1954dimensionless}
\bibinfo{author}{A.~Monin}, \bibinfo{author}{A.~Obukhov},
\newblock \bibinfo{title}{Dimensionless characteristics of turbulence in the
  surface layer},
\newblock \bibinfo{journal}{Akad. Nauk SSSR, Geofiz. Inst., Tr}
  \bibinfo{volume}{24} (\bibinfo{year}{1954}) \bibinfo{pages}{163--187}.
\bibitem[{Hellmann(1919)}]{hellmann1919bewegung}
\bibinfo{author}{G.~Hellmann},
\newblock \bibinfo{title}{{\"U}ber die bewegung der luft in den untersten
  schichten der atmosph{\"a}re, dritte mitteilung},
\newblock \bibinfo{journal}{Sitzungsber. Akad. d. Wiss. Berlin}
  (\bibinfo{year}{1919}) \bibinfo{pages}{404--416}.
\bibitem[{Smedman-H{\"o}gstr{\"o}m and
  H{\"o}gstr{\"o}m(1978)}]{smedman1978practical}
\bibinfo{author}{A.-S. Smedman-H{\"o}gstr{\"o}m},
  \bibinfo{author}{U.~H{\"o}gstr{\"o}m},
\newblock \bibinfo{title}{A practical method for determining wind frequency
  distributions for the lowest 200 m from routine meteorological data},
\newblock \bibinfo{journal}{Journal of Applied Meteorology and Climatology}
  \bibinfo{volume}{17} (\bibinfo{year}{1978}) \bibinfo{pages}{942--954}.
\bibitem[{Panofsky et~al.(1960)Panofsky, Blackadar, and
  McVehil}]{panofsky1960diabatic}
\bibinfo{author}{H.~Panofsky}, \bibinfo{author}{A.~Blackadar},
  \bibinfo{author}{G.~McVehil},
\newblock \bibinfo{title}{The diabatic wind profile},
\newblock \bibinfo{journal}{Quarterly Journal of the Royal Meteorological
  Society} \bibinfo{volume}{86} (\bibinfo{year}{1960})
  \bibinfo{pages}{390--398}.
\bibitem[{Hussain and Dutilleux(2020)}]{hussain2020parametric}
\bibinfo{author}{H.~Hussain}, \bibinfo{author}{G.~Dutilleux},
\newblock \bibinfo{title}{A parametric study of long-range atmospheric sound
  propagation using underwater acoustics software},
\newblock in: \bibinfo{booktitle}{Proceedings of Meetings on Acoustics LRSP},
  volume~\bibinfo{volume}{41}, \bibinfo{organization}{Acoustical Society of
  America}, \bibinfo{year}{2020}, p. \bibinfo{pages}{022001}.
\bibitem[{Hansen et~al.(2019)Hansen, Zajam{\v{s}}ek, and
  Hansen}]{hansen2019investigation}
\bibinfo{author}{K.~L. Hansen}, \bibinfo{author}{B.~Zajam{\v{s}}ek},
  \bibinfo{author}{C.~H. Hansen},
\newblock \bibinfo{title}{Investigation of a microphone height correction for
  long-range wind farm noise measurements},
\newblock \bibinfo{journal}{Applied Acoustics} \bibinfo{volume}{155}
  (\bibinfo{year}{2019}) \bibinfo{pages}{97--110}.
\bibitem[{Salomons(2001)}]{salomons2001computational}
\bibinfo{author}{E.~Salomons}, \bibinfo{title}{Computational atmospheric
  acoustics}, \bibinfo{publisher}{Springer Science \& Business Media},
  \bibinfo{year}{2001}.
\bibitem[{Saltelli et~al.(2010)Saltelli, Annoni, Azzini, Campolongo, Ratto, and
  Tarantola}]{saltelli2010variance}
\bibinfo{author}{A.~Saltelli}, \bibinfo{author}{P.~Annoni},
  \bibinfo{author}{I.~Azzini}, \bibinfo{author}{F.~Campolongo},
  \bibinfo{author}{M.~Ratto}, \bibinfo{author}{S.~Tarantola},
\newblock \bibinfo{title}{Variance based sensitivity analysis of model output.
  design and estimator for the total sensitivity index},
\newblock \bibinfo{journal}{Computer physics communications}
  \bibinfo{volume}{181} (\bibinfo{year}{2010}) \bibinfo{pages}{259--270}.
\bibitem[{Yip et~al.(2011)Yip, Ferro, Stephenson, and Hawkins}]{yip2011simple}
\bibinfo{author}{S.~Yip}, \bibinfo{author}{C.~A. Ferro}, \bibinfo{author}{D.~B.
  Stephenson}, \bibinfo{author}{E.~Hawkins},
\newblock \bibinfo{title}{A simple, coherent framework for partitioning
  uncertainty in climate predictions},
\newblock \bibinfo{journal}{Journal of Climate} \bibinfo{volume}{24}
  (\bibinfo{year}{2011}) \bibinfo{pages}{4634--4643}.
\bibitem[{Attenborough(1985)}]{attenborough1985acoustical}
\bibinfo{author}{K.~Attenborough},
\newblock \bibinfo{title}{Acoustical impedance models for outdoor ground
  surfaces},
\newblock \bibinfo{journal}{Journal of Sound and Vibration}
  \bibinfo{volume}{99} (\bibinfo{year}{1985}) \bibinfo{pages}{521--544}.
\bibitem[{Attenborough et~al.(2002)Attenborough, Bashir, and
  Taherzadeh}]{attenborough2002review}
\bibinfo{author}{K.~Attenborough}, \bibinfo{author}{I.~Bashir},
  \bibinfo{author}{S.~Taherzadeh},
\newblock \bibinfo{title}{A review of ground impedance models for propagation
  modelling},
\newblock \bibinfo{journal}{University of Hull, Hull, UK, Tech. Rep}
  (\bibinfo{year}{2002}).
\bibitem[{Nguyen et~al.(2021)Nguyen, Hansen, Catcheside, Hansen, and
  Zajamsek}]{nguyen2021long}
\bibinfo{author}{P.~D. Nguyen}, \bibinfo{author}{K.~L. Hansen},
  \bibinfo{author}{P.~Catcheside}, \bibinfo{author}{C.~Hansen},
  \bibinfo{author}{B.~Zajamsek},
\newblock \bibinfo{title}{Long-term quantification and characterisation of wind
  farm noise amplitude modulation},
\newblock \bibinfo{journal}{Measurement}  (\bibinfo{year}{2021})
  \bibinfo{pages}{109678}.
\bibitem[{9613-2(1996)}]{ISO9613}
\bibinfo{author}{I.~9613-2}, \bibinfo{title}{Acoustics: Attenuation of sound
  during propagation outdoors}, \bibinfo{year}{1996}.
\bibitem[{Komatsu(2008)}]{komatsu2008improvement}
\bibinfo{author}{T.~Komatsu},
\newblock \bibinfo{title}{Improvement of the delany-bazley and miki models for
  fibrous sound-absorbing materials},
\newblock \bibinfo{journal}{Acoustical science and technology}
  \bibinfo{volume}{29} (\bibinfo{year}{2008}) \bibinfo{pages}{121--129}.
\bibitem[{Plovsing and Kragh(2006)}]{plovsing1849nord2000}
\bibinfo{author}{B.~Plovsing}, \bibinfo{author}{J.~Kragh},
\newblock \bibinfo{title}{Nord2000. comprehensive outdoor sound propagation
  model. part 1: Propagation in an atmosphere without significant refraction},
\newblock \bibinfo{journal}{DELTA Acoustics \& Vibration, Report AV}
  (\bibinfo{year}{2006}).
\bibitem[{Lihoreau et~al.(2006)Lihoreau, Gauvreau, B{\'e}rengier, Blanc-Benon,
  and Calmet}]{lihoreau2006outdoor}
\bibinfo{author}{B.~Lihoreau}, \bibinfo{author}{B.~Gauvreau},
  \bibinfo{author}{M.~B{\'e}rengier}, \bibinfo{author}{P.~Blanc-Benon},
  \bibinfo{author}{I.~Calmet},
\newblock \bibinfo{title}{Outdoor sound propagation modeling in realistic
  environments: Application of coupled parabolic and atmospheric models},
\newblock \bibinfo{journal}{The Journal of the Acoustical Society of America}
  \bibinfo{volume}{120} (\bibinfo{year}{2006}) \bibinfo{pages}{110--119}.
\bibitem[{Salomons(1994)}]{salomons1994downwind}
\bibinfo{author}{E.~M. Salomons},
\newblock \bibinfo{title}{Downwind propagation of sound in an atmosphere with a
  realistic sound-speed profile: A semianalytical ray model},
\newblock \bibinfo{journal}{The Journal of the Acoustical Society of America}
  \bibinfo{volume}{95} (\bibinfo{year}{1994}) \bibinfo{pages}{2425--2436}.
\bibitem[{Heimann and Gross(1999)}]{heimann1999coupled}
\bibinfo{author}{D.~Heimann}, \bibinfo{author}{G.~Gross},
\newblock \bibinfo{title}{Coupled simulation of meteorological parameters and
  sound level in a narrow valley},
\newblock \bibinfo{journal}{Applied Acoustics} \bibinfo{volume}{56}
  (\bibinfo{year}{1999}) \bibinfo{pages}{73--100}.
\bibitem[{Mo et~al.(2017)Mo, Yeh, Lin, and Manocha}]{mo2017outdoor}
\bibinfo{author}{Q.~Mo}, \bibinfo{author}{H.~Yeh}, \bibinfo{author}{M.~Lin},
  \bibinfo{author}{D.~Manocha},
\newblock \bibinfo{title}{Outdoor sound propagation with analytic ray curve
  tracer and gaussian beam},
\newblock \bibinfo{journal}{The Journal of the Acoustical Society of America}
  \bibinfo{volume}{141} (\bibinfo{year}{2017}) \bibinfo{pages}{2289--2299}.

\end{thebibliography}

\end{document}